\documentclass[
    aps,
    reprint,
    prb,
	twocolumn,
	showpacs,
	preprintnumbers,
	amsmath,
	amssymb]{revtex4-1}

\usepackage{graphicx}
\usepackage{upgreek}
\usepackage{color}
\usepackage{xspace}
\usepackage{textcomp}
\usepackage{upgreek}
\usepackage{siunitx}
\usepackage{placeins}

\newcommand{\perovskite}[2]{#1#2O$_3$\xspace}
\newcommand{\BBO}{\perovskite{Ba}{Bi}}
\newcommand{\BPO}{\perovskite{Ba}{Pb}}
\newcommand{\STO}{\perovskite{Sr}{Ti}}

\newcommand{\eg}{e.\,g.\xspace}
\newcommand{\ie}{i.\,e.\xspace}
\newcommand{\Tc}{\ensuremath{T_\mathrm c}\xspace}
\newcommand{\RH}{\ensuremath{R_\mathrm H}\xspace}

\newcommand{\Tref}[1][]{\ensuremath{T_\mathrm{ref{#1}}}\xspace}
\newcommand{\tauso}[1][]{\ensuremath{\tau_\mathrm{so{#1}}}\xspace}
\newcommand{\tauo}[1][]{\ensuremath{\tau_\mathrm{o{#1}}}\xspace}
\newcommand{\taui}[1][]{\ensuremath{\tau_\mathrm{i{#1}}}\xspace}
\newcommand{\Bso}[1][]{\ensuremath{B_\mathrm{so{#1}}}\xspace}
\newcommand{\Bo}[1][]{\ensuremath{B_\mathrm{o{#1}}}\xspace}
\newcommand{\Bi}[1][]{\ensuremath{B_\mathrm{i{#1}}}\xspace}
\newcommand{\dBPO}[1][]{\ensuremath{d_\mathrm{BPO}}\xspace}
\newcommand{\BiSe}{Bi$_2$Se$_3$\xspace}

\begin{document}

\title{Electronic transport in thin films of \BPO: Unraveling two-dimensional quantum effects}

\author{P.~Seiler}
\author{R.~Bartel}
\author{T.~Kopp}
\author{G.~Hammerl}
\email{german.hammerl@physik.uni-augsburg.de} 
\affiliation{Center for Electronic Correlations and Magnetism, Experimental Physics VI, Institute of Physics, University of Augsburg, 86135 Augsburg, Germany}

\date{\today}

\begin{abstract}
Recently, perovskite related \BPO has attracted attention due to its hidden topological properties and, 
moreover, has been used as a thin layer in heterostructures to induce two-dimensional superconductivity.
Here we investigate the normal state electronic transport properties of thin films of \BPO. 
Temperature and magnetic field dependent sheet resistances are strongly affected by two-dimensional quantum effects. 
Our analysis decodes the interplay of spin--orbit coupling, disorder, and electron--electron interaction in this compound. 
Similar to recently discussed topological materials, we find that weak antilocalization is the dominant protagonist in 
magnetotransport, whereas electron--electron interactions play a pronounced role in the temperature dependence.
A systematic understanding of these quantum effects is essential to allow for an accurate control of properties 
not only of thin films of \BPO, but also of topological heterostructures.
\end{abstract}

\maketitle

\section{Introduction}

Oxide heterostructures serve as a new source of functionality by 
combining intrinsic physical properties of different compounds in epitaxially 
grown artificial materials~\cite{tsymbal:2012,freericks:2016}.
Novel intriguing qualities emerge from the  tunability of coupled degrees of freedom like charge,
spin and orbital degrees, or by reduced dimensionality~\cite{zubko:2011}. 
Oxide heterostructures with their interfaces can do even more to realize 
multi-functional devices: as compared to, \eg, semiconductor heterostructures, 
they allow for emergent electronic phases that are not stable in the bulk. 
Such electronic phases can be well constrained on the nanoscale, 
often being correlated electronic systems driven by 
electronic reconstruction~\cite{hesper:2000,okamoto:2004,hwang:2012}. 
Moreover, the inversion symmetry is generically broken 
at these interfaces, leading to a strong Rashba-type spin--orbit coupling~\cite{rashba:1960,Hikami:1980} 
and inducing exotic spin textures or even superconducting states that are characterized by non-trivial values of topological invariants~\cite{sau:2010,loder:2015}. 
Therefore, transport in these two-dimensional systems is intrinsically intriguing and quantum corrections have an immense impact on transport properties at low temperatures. 
Self-interference of extended electron waves or electron--electron interaction effects decide on metallic or insulating ground states.
Specifically, disorder in two-dimensional systems causes weak localization~\cite{Hikami:1980, Altshuler:1980a, Lee:1985} and an 
insulating ground state~\cite{Abrahams:1979, Wegner:1979}, whereas its antagonist, weak antilocalization~\cite{Hikami:1980, Wegner:1989, Iordanskii:1994},
originating from a combination of disorder and spin--orbit coupling, results in a metallic ground state.
In this context of quantum transport, the two perovskite-related compounds \BBO and \BPO are promising candidates to be studied:
both are expected to preserve a ``hidden'' topological insulator phase when electron or hole doped~\cite{yan:2013,li:2015}. 
In \BPO--\BBO bilayers grown on \STO substrates we recently observed two-dimensional superconductivity with maximum \Tc of $\SI{3.5}{\kelvin}$,
presumably induced by interfacial strain where \BPO is acting as a dopant layer~\cite{meir:2017}. 
This is astounding as \BBO is a charge-density wave ordered insulator~\cite{cox:1976,cox:1979,varma:1988},
whereas \BPO a metal which becomes superconducting only at temperatures below $\SI{0.5}{\kelvin}$~\cite{bogato:1980}.

To gain a better control and understanding of heterostructures involving \BPO, we study the transport properties of a thin layer of \BPO  on top of a \STO substrate. 
Due to the strong spin--orbit coupling in \BPO layers, we expect a sizable weak antilocalization contribution in our samples at low temperatures. 
In this article, we show experimental data retrieved from magnetotransport measurements alongside the temperature dependence of the sheet resistance. 
Whereas metallic features of weak antilocalization dominate the magnetic field dependence, insulating features of electron--electron interaction predominate in 
controlling the temperature dependence of the resistance. 
It turns out that a thoughtful data analysis is necessary to explain the---at first sight contradicting---experimental results. 

\section{Methods}
\label{sec:meths}

\subsection{Sample Growth}

All thin films of \BPO were grown by pulsed laser deposition (PLD) using commercially available 
stoichiometric targets with purities of at least 99.95\% at maximum achievable density. 
The surface of the targets were cleaned for each sample growth to maintain the quality of 
the samples. 
We used single crystalline $\SI{5}{\milli\metre}\times\SI{5}{\milli\metre}\times\SI{1}{\milli\metre}$ 
sized (001) oriented one-side polished \STO crystals as substrates the surface of which were
HF treated~\cite{kawasaki:1994,koster:1998} and annealed in oxygen in order to guarantee extensive TiO$_2$ termination. 
The substrates were fixed by silver paste on heating platforms before being transferred to the vacuum chamber. 
The substrates were heated to $\approx\SI{552}{\celsius}$ either by resistive or laser heating controlled 
by pyrometers in about $\SI{45}{\minute}$ at an oxygen background pressure of at least $\SI{0.25}{\milli\bar}$.
Our PLD system is equipped with a KrF excimer laser having a nominal fluency of $\SI{2}{\joule\per\square\centi\metre}$. 
During thin film growth of \BPO the oxygen background pressure was leveled to $\approx\SI{1}{\milli\bar}$. 
The laser pulse energy was set to $\SI{550}{\milli\joule}$ for all samples and the laser shot frequency was limited 
to~$\SI{3}{\hertz}$. After thin film deposition the samples were cooled to $\SI{400}{\celsius}$ within~$\SI{3}{\minute}$ 
and annealed at an oxygen background pressure of $\approx\SI{400}{\milli\bar}$ for at least~$\SI{20}{\minute}$ 
before they freely cooled to room temperature.

\subsection{Sample Preparation}

The topography of the samples was routinely checked by atomic force microscopy (AFM). 
The grown \BPO thin films only display (00$l$) oriented peaks in X-ray diffraction (XRD) experiments. 
We used X-ray reflectometry (XRR) to check the film thicknesses of the samples, where we found the expected 
linear dependence of the film thickness in respect to the number of laser pulses used during film synthesis. 
The samples were patterned using standard photolithography and argon ion-etching resulting in measurement bars of about $\SI{200}{\micro\metre}$ 
in length and $\SI{50}{\micro\metre}$ in width.
Transport properties were measured in a commercial 14\,T PPMS in common four-point geometry.

\section{Impact of Quantum Corrections}
\label{sec:Theory}

\subsection{Quantum Interference}

Due to the strong atomic spin--orbit coupling of lead, as well as the broken inversion symmetry 
of the interface, D'yakonov--Perel spin relaxation~\cite{dyakonov:1971} is expected to affect transport in \BPO thin films. 
In combination with disorder, the spin relaxation results in a signature of weak antilocalization (WAL)~\cite{Bergmann:1984,Nakamura:2012,Kim:2013,Seiler:2018,Pai:2018},
predicting a logarithmic decrease of the resistance upon cooling.
In the presence of time-reversal symmetry breaking magnetic fields, the same correction causes a positive magnetoresistance
that allows to extract the strengths of inelastic scattering as well as spin relaxation. 
An adequate description of the first order quantum correction to the conductivity of a two-dimensional system is 
given by the Iordanskii--Lyanda-Geller--Pikus theory~\cite{Iordanskii:1994,Pikus:1995,Knap:1996}. 
The specific magnetic field dependence is sensitive to the winding number of the spin expectation value traced 
along the Fermi surface. 
In the case of a triple spin winding, which has been identified in several \STO 
based two-dimensional systems~\cite{Nakamura:2012, Seiler:2018, Pai:2018} and is also in good agreement with our data,
the Iordanskii--Lyanda-Geller--Pikus theory reproduces the analytical result of the Hikami--Nagaoka--Larkin theory~\cite{Hikami:1980}.
The first order quantum correction to the conductivity in magnetic field~$B$ due to quantum interference (QI) is given by
\begin{align}
\updelta \sigma^\text{QI} (B) &= \frac{e^2}{\pi h} 
\left[\uppsi\left(\frac{1}{2}+\frac{\Bso+\Bi}{B}\right)
-\frac{1}{2}\uppsi\left(\frac{1}{2}+\frac{\Bi}{B}\right)\nonumber\right.\\[0.2cm]
&\left.+
\frac{1}{2}\uppsi\left( \frac{1}{2} + \frac{2\Bso+\Bi}{B}\right)
-\uppsi\left(\frac{1}{2}+\frac{\Bo}{B}\right)
\right],
\label{eq:Hikami}
\end{align}
with $\uppsi$ being the digamma function.
The effective magnetic fields are defined by
\begin{equation}
\Bso[/i/o] = \frac{\hbar}{4eD\tauso[/i/o]},
\end{equation}
with $D$ being the diffusion constant, $\tauo$ and $\taui$ the elastic and inelastic scattering 
times and $\tauso$ the time scale associated with the D'yakonov--Perel spin relaxation~\cite{dyakonov:1972}.

For vanishing magnetic field, there are two relevant limits of Eq.~\eqref{eq:Hikami}: 
$\Bso \gg \Bi$, which relates to a metallic regime, 
and $\Bso \ll \Bi$, which corresponds to an insulating regime~\cite{Hikami:1980, Wegner:1989}. 
The correction describing the metallic regime is given by
\begin{equation}
\updelta \sigma^\text{WAL} (B \to 0) = 
-\frac{e^2}{2 \pi h} \ln \left( \frac{\Bo^2 \Bi}{2\Bso^3 } \right).
\end{equation}
By assuming an algebraic temperature dependence, 
\begin{equation}
\Bi = \gamma + \beta T^\alpha,  
\label{eq:bitemp}
\end{equation}
with temperature exponent $\alpha$, the WAL quantum correction takes the form
\begin{equation}
\updelta \sigma^\text{WAL} (T) = - \frac{ e^2}{2 \pi h} \ln \left( 
\frac{\gamma + \beta T^\alpha}{C^\text{WAL}} \right),
\label{eq:sigmaTsym}
\end{equation}
where $C^\text{WAL}$ is a temperature independent constant. 
Equation~\eqref{eq:sigmaTsym} describes an increasing conductivity for decreasing temperature, indicating a metallic ground state.
Note that the inelastic field does not necessarily vanish for zero temperature, due to a finite constant~$\gamma$ in Eq.~\eqref{eq:sigmaTsym}.
This has been observed in several magnetotransport measurements~\cite{Mohanty1997a, Mohanty1997b, Imry2008}.
Also mind that the effect of spin--orbit coupling enters the conductivity correction in Eq.~\eqref{eq:sigmaTsym} only via the constant $C^\text{WAL}$,
whereas the inelastic scattering processes are essentially responsible for the temperature dependence.

For the insulating case of weak localization (WL; $\Bso\ll\Bi)$, Eq.~\eqref{eq:Hikami} reduces to
\begin{equation}
\updelta \sigma^\text{WL} (B \to 0) =  \frac{e^2}{ \pi h} \ln \left( \frac{\Bi}{\Bo} \right),
\end{equation}
and in terms of a temperature dependent correction to the conductivity,
\begin{equation}
\updelta \sigma^\text{WL} (T) = \frac{e^2}{ \pi h} \ln \left( \frac{\gamma + \beta T^\alpha}{C^\text{WL}}  \right).
\label{eq:sigmaTort}
\end{equation}
Equation~\eqref{eq:sigmaTort} describes a decrease of conductivity when the temperature is lowered and is 
the well-known logarithmic correction of WL.

We refer to the magnetoconductivity in magnetic fields $B \ll \Bo$ as
\begin{align}
&\Delta\sigma^\text{QI} (B) = \updelta \sigma^\text{QI}(B) - \updelta \sigma^\text{QI}(0) \label{eq:Hikami2} \\[0.2cm] 
&= \frac{e^2}{\pi h} \left[\Psi\left( \frac{B}{\Bso + \Bi} \right) 
-\frac{1}{2} \Psi\left( \frac{B}{\Bi} \right) \nonumber \right. \left.
+\frac{1}{2} \Psi\left(  \frac{B}{2\Bso+\Bi}\right)
\right],
\end{align}
where $\Psi(x)=\ln(x)+\uppsi(\tfrac{1}{2}+\tfrac{1}{x})$. 
Note that in the metallic case $(\Bso \gg \Bi)$, the conductivity up to quadratic order in the magnetic field
is again driven by the inelastic scattering field, whereas the spin--orbit coupling fixes the prefactor, 
\begin{equation}
\Delta\sigma^\text{WAL} (B) \approx - \frac{e^2}{\pi h} \frac{1}{48} \left( \frac{B}{\Bi} \right)^2.
\label{eq:WALsmall}
\end{equation}

As the magnetic field and temperature dependence of the conductivity originates from the same analytic 
expression of the quantum correction, we emphasize that the appearance of WAL in terms of a positive 
magnetoresistance (following Eq.~\eqref{eq:WALsmall}) is intrinsically connected to a logarithmic decrease of the sheet resistance for lower temperature (following Eq.~\eqref{eq:sigmaTsym}). 

\subsection{Electron--Electron Interaction}

A logarithmic increase in the sheet resistance for lower temperatures, however, cannot only be caused by WL, but also by electron--electron interaction (EEI) in the particle--hole channel.
The corresponding correction to the conductivity in two dimensions is given by~\cite{Altshuler:1980a, Altshuler:1980b, Lee:1985}
\begin{equation}
\updelta \sigma^\text{EEI}(T) = \zeta\frac{e^2}{\pi h}  \ln \left( \frac{T}{C^\text{EEI}} \right),
\label{eq:qceeiT}
\end{equation}
where~$\zeta$ is in the range between $\zeta=1$ without screening and $\zeta \sim 0.35$ for perfect screening, and $C^\text{EEI}$ is a temperature independent constant.  
Although the particle--hole channel is not sensitive to the magnetic field directly, a finite Zeeman splitting produces a magnetoconductivity following
\begin{equation}
\Delta \sigma^\text{EEI} (\tilde B) \approx -\frac{e^2}{ \pi h} \frac{2 \left( 1 - \zeta \right)}{3}  \begin{cases}
\ln \left( \frac{\tilde B }{1.3 } \right)  & \tilde B \gg 1 \\[0.3cm]
0.084 \tilde B^2 & \tilde B \ll 1 
\end{cases}
\label{eq:qceei}
\end{equation}
for~$\tilde B=g \mu_\text{B} B / k_\text{B} T$. 
Therefore, magnetoresistance due to EEI is expected to grow monotonously with larger fields. This is in contrast with the decreasing magnetoresistance for higher magnetic fields in the case of WAL and, even more, with the
negative progression of magnetoresistance for all magnetic fields in case of WL.

\section{Experimental Results and Discussion}
\label{sec:exp}

Magnetotransport data taken on various \BPO thin film samples show no indication of multi-band behavior (see~Appendix~\ref{ap:hall}).
This is an advantage over the more complicated evaluation of multi-band systems, where not only the Hall effect~\cite{Seiler:2018},
but also supervening aspects of quantum corrections have to be considered~\cite{Seiler:2019,seiler:2019a}. 
The logarithmic dependences we find in the progression of the resistance towards lower temperatures
are analyzed and quantitatively described by the quantum corrections discussed in Sec.~\ref{sec:Theory}.
In the following, we show data obtained on a $\dBPO=\SI{4.8}{\nano\metre}$ thick \BPO thin film sample acting as a typical representative.
In Appendix~\ref{ap:sample}, the same data analysis is shown for a thicker sample with $\dBPO=\SI{21.3}{\nano\metre}$, confirming our results.

\subsection{Analysis of Magnetoresistance}
\label{subsec:MR}

\begin{figure}
\centering
\includegraphics[width=\columnwidth]{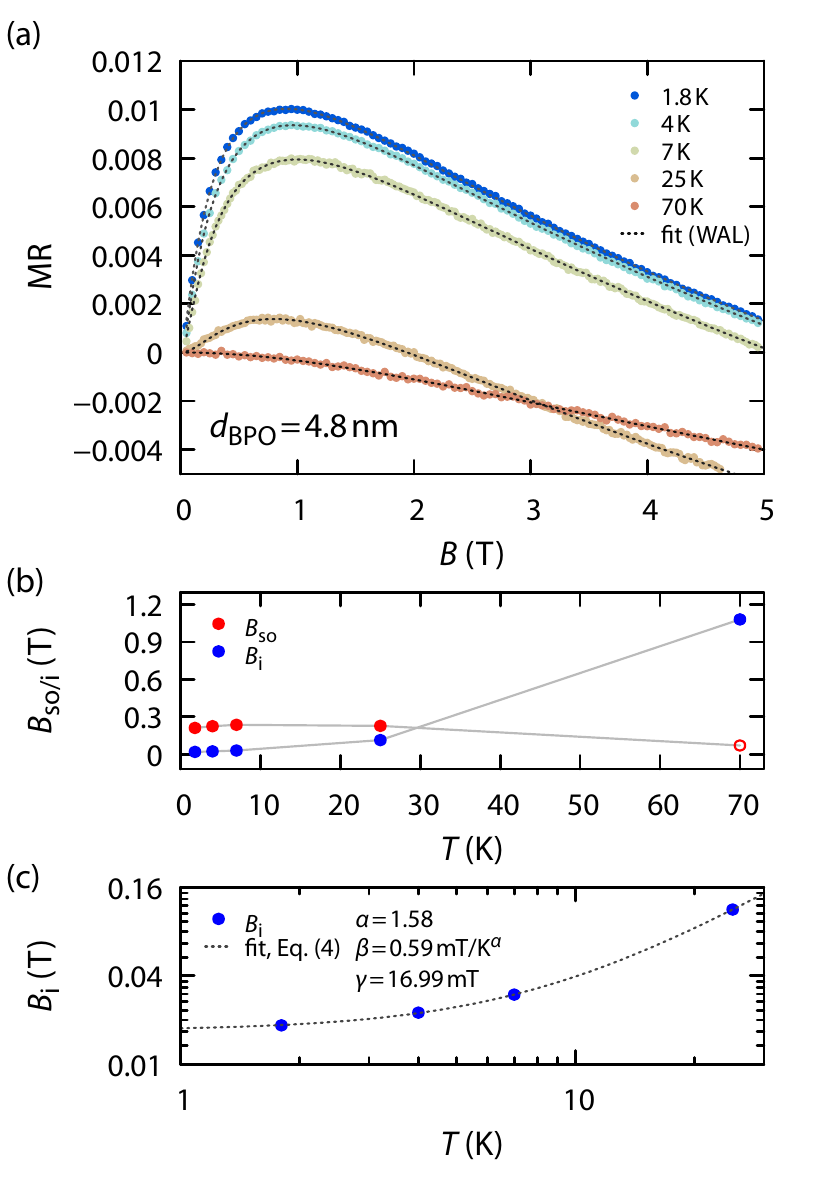}
\caption{Exemplary magnetoresistance (MR) data and fit parameters taken and retrieved from a $\SI{4.8}{\nano\metre}$ thick \BPO thin film
grown on a (001) oriented \STO substrate. 
(a) Temperature-dependent magnetoresistance as function of an applied external magnetic field. 
The dotted lines are fits modeling the quantum corrections following Eq.~\eqref{eq:Hikami2}. 
(b) Extracted spin--orbit fields \Bso (red) and inelastic fields \Bi (blue). 
At low temperatures \Bso is clearly larger than \Bi with maximum values reaching $\SI{0.24}{\tesla}$.
At high temperatures $\Bi\gg\Bso$, therefore the fitted value of \Bso looses its significance (red circle). 
Gray lines are guides to the eye.
(c) Temperature dependence of the inelastic magnetic field \Bi following Eq.~\eqref{eq:bitemp}.
} 
\label{fig:wal5nm}
\end{figure}

The magnetoresistance data are positive for small magnetic fields and display a distinct WAL maximum at $B \sim 0.8$\,T, see~Fig.~\ref{fig:wal5nm}\,(a). 
Fits using Eq.~\eqref{eq:Hikami2} describe the data over the full measured magnetic field range, indicating a triple spin winding.
The maximum value of the extracted spin--orbit field is given by $\Bso=\SI{0.24}{\tesla}$. 

The second fitting parameter is the inelastic scattering field \Bi. 
At temperatures as high as \SI{70}{\kelvin}, \Bi clearly exceeds \Bso, accessing the WL regime, see Fig.~\ref{fig:wal5nm}\,(b). 
Here the magnetoresistance is steadily decreasing for all magnetic fields.
Below, the WAL quantum correction can be extracted for temperatures up to 25\,K, which we use as a reference temperature \Tref in the following analysis.
We estimate the temperature dependence of \Bi up to \Tref to be described by an exponent~$\alpha = 1.58$ as well as by a constant contribution for zero temperature, 
$\gamma =\SI{17}{\milli\tesla}$, see Fig.~\ref{fig:wal5nm}\,(c). 
The exponent we find is in the typical range for two-dimensional systems, located in the crossover regime between the linear temperature dependence ($\alpha=1$) due to dephasing by electron-electron scattering at very low temperatures and due to dephasing by electron-phonon scattering ($\alpha\geq2$) at slightly higher temperatures~\cite{Abrahams:1981,Lin:2002}.
Note that the matching of our data with Eq.~\eqref{eq:Hikami2} is a strong indication for single channel transport~\cite{Lu2014,Liao:2017}.

Before proceeding with the analysis of the temperature dependence of the sheet resistance, we like to discuss the unambiguousness of our WAL analysis. 
Positive magnetoresistance typically results from several origins: 
(a) multi-carrier Hall effect,
(b) magnetism,
(c) superconducting fluctuations~\cite{Aslamasov:1968, Maki:1968, Thompson:1970},
(d) EEI~\cite{Altshuler:1980a, Altshuler:1980b, Lee:1985}, or
(e) WAL. 

A multi-carrier Hall effect 
(a) does not only affect the sheet resistance, but also most prominently the Hall measurement in form of non-linear contributions.
However, as already addressed above, the Hall signal is linear in all of our samples even up to 14\,T (see~Appendix~\ref{ap:hall}) and a multi-carrier effect can be excluded. 
There is also no obvious signature of magnetism in our samples 
(b), \eg, no hysteresis is observed in  magnetotransport. 
In addition, we see no indications for superconductivity even at temperatures as low as~1.8\,K, therefore an imprint of superconducting fluctuations 
(c) is hardly probable. 
It remains to differentiate the impact of EEI (d) versus WAL (e) on magnetotransport.

The appearance of sizable EEI has been reported in several oxide heterostructures, in particular along with positive magnetoresistance~\cite{Fuchs:2015}.
However, this quantum correction provides a distinct magnetic field behavior in two dimensions:
the increase of resistance is quadratic and positive in small magnetic fields and logarithmically increasing for higher fields, see Eq.~\eqref{eq:qceei}.
The observed distinct maximum with a decreasing or even negative resistance for higher magnetic fields cannot be explained in the scope of EEI.

A combination of WL and EEI quantum corrections has been discussed in two-dimensional semiconductor structures and metallic thin films~\cite{poole:1981,bishop:1982,lin:1984,bergmann:1984a,taboryski:1990}.
Pronounced EEI together with a rather suppressed WL contribution could in principle resemble the observed MR progression.
However, our MR data cannot be fitted with physically meaningful parameters by
using a combination of WL and EEI quantum corrections.

Any interaction contribution in the magnetotransport data must therefore be much smaller than the quantum interference contribution.
Furthermore, the amplitude of the observed magnetoresistance is rather large: the WAL fits provide values of the inelastic scattering field of 
$\Bi \sim \SI{20}{\milli\tesla}$ for low temperatures (see Fig.~\ref{fig:wal5nm}).
Considering Eq.~\eqref{eq:WALsmall}, this corresponds to a prefactor that is several orders of magnitudes higher than any realistic prefactor in Eq.~\eqref{eq:qceei},
where $\tilde{B}\sim 1$ for the relevant temperature regime.

Therefore, WAL (e) is the dominant effect that is in agreement with our magnetotransport data,
reinforced by the perfect viability of the WAL fits with their realistic fitting parameters.

\subsection{Analysis of Sheet Resistance} 
\label{subsec:RT}

\begin{figure}
\centering
\includegraphics[width=\columnwidth]{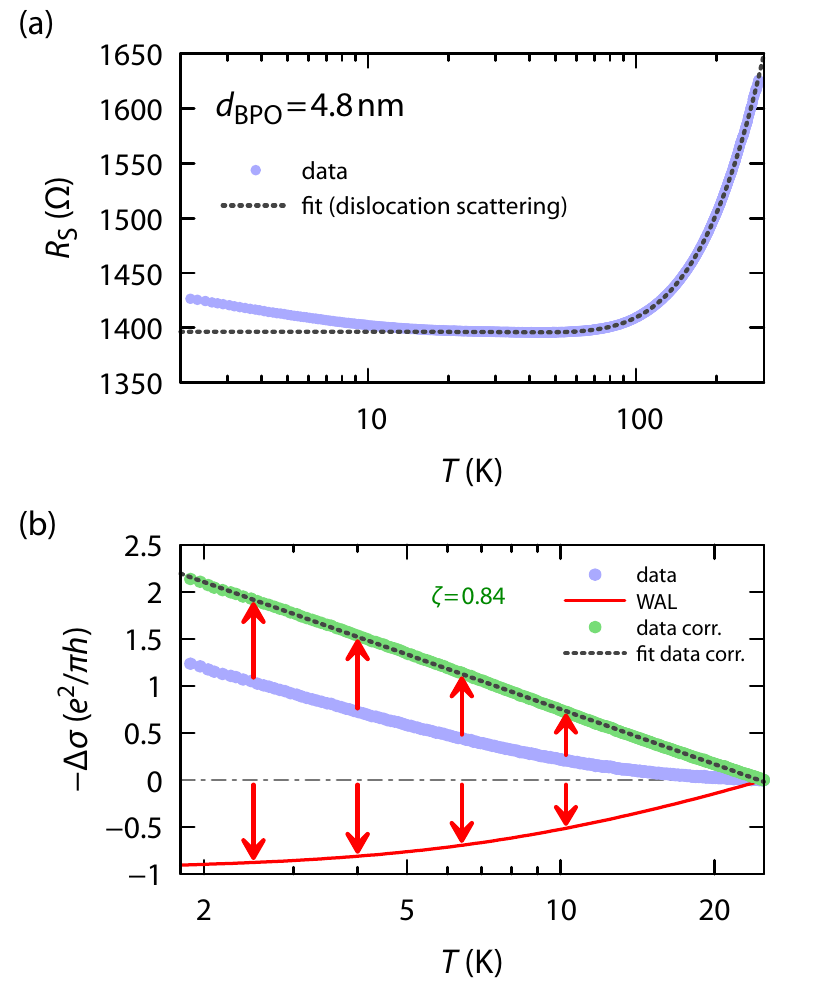}
\caption{(a) Observed dependence of the sheet resistance~($R_\text{S}$) on temperature.
The progression can be understood in terms of electron--phonon scattering at high temperatures,
but is also dominated by dislocation scattering~\cite{Fuchs:2015}. 
(b) The progression of the change in conductivity in respect to the reference temperature $\Tref=\SI{25}{\kelvin}$ in logarithmic scale (blue data).
Following the WAL analysis, a metallic ground state is expected (red line). Reevaluating the data by subtraction of the WAL contribution (red arrows)
reveals a logarithmic progression (green data corr.). 
The fit of this progression is in excellent agreement with weakly screened EEI.}
\label{fig:rt-5nm}
\end{figure}

Now we turn to the temperature dependence of the resistance.  
The sheet resistance shows a minimum for an intermediate temperature regime (around~\SI{50}{\kelvin}) and increases for decreasing temperature (see Fig.~\ref{fig:rt-5nm}).
The minimum can be well explained by thermally activated dislocation scattering~\cite{Fuchs:2015}.
An additional sheet resistance increase for lower temperatures, however, is not anticipated from the magnetoresistance analysis.
The WAL in the regime $\Bso \gg \Bi$ for temperatures below $\Tref=\SI{25}{\kelvin}$ is associated with a quantum correction that describes a decrease of resistance jointly with temperature, raising the expectation for a metallic ground state.
Furthermore, as we observe only one conduction channel (see Appendix~\ref{ap:hall}), a simultaneous occurrence of WL accompanied by WAL is hardly probable.

\begin{figure}
\centering
\includegraphics[width=\columnwidth]{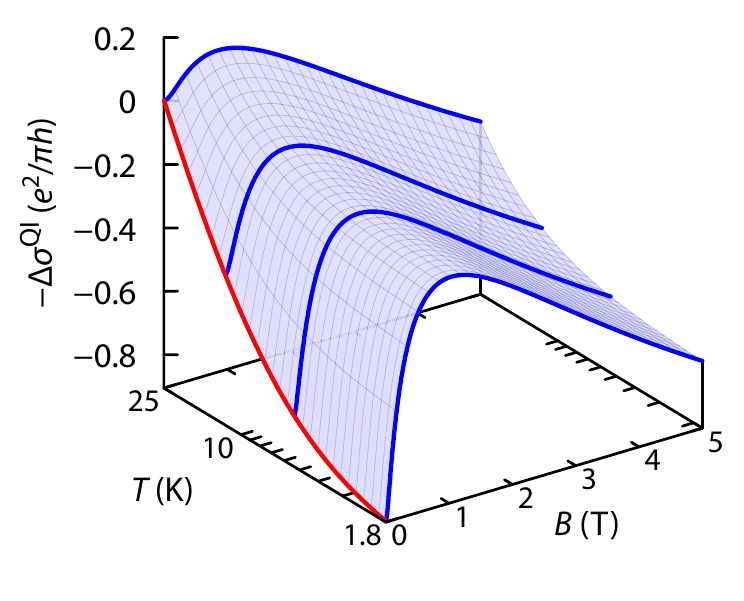}
\caption{Progression of the negative conductivity correction over temperature $T$ and magnetic field $B$ following~Eq.~\eqref{eq:3d}
in respect to the chosen reference temperature $\Tref=\SI{25}{\kelvin}$ and setting $\Bso=\SI{0.24}{\tesla}$ as extracted from the fits.
The blue lines indicate where data are taken in magnetoresistance measurements.
We extrapolate the magnetoconductivity to the case $B=0$ (red line).
This is the corresponding WAL quantum correction for zero magnetic field.}
\label{fig:wal3d}
\end{figure}

As shown in Sec.~\ref{sec:Theory}, the magnetoresistance for fixed temperature and the temperature-dependent resistance for fixed field (\ie zero magnetic field) are described by the same analytical expression. 
As we have excluded further effects contributing to the magnetoresistance, we extrapolate the quantum correction, 
\begin{equation}
\label{eq:3d}
\begin{split}
\Delta\sigma^{\mathrm{QI}}(B,\Bi(T),\Bso)  & =   \\
\updelta\sigma^{\mathrm{QI}}\left(B,\Bi(T),\Bso\right) & -\updelta\sigma^{\mathrm{QI}}\left(0,\Bi(\Tref),\Bso\right),
\end{split}
\end{equation}
to zero magnetic field, see Fig.~\ref{fig:wal3d}. 
Here we use the spin--orbit field $\Bso=\SI{0.25}{\tesla}$ as extracted from MR data and $\Tref=\SI{25}{\kelvin}$ as the reference temperature.
This extrapolated WAL contribution is then subtracted from the measured temperature dependent resistances. 
With this adjustment, a positive logarithmic temperature dependence at low temperatures is uncovered, see Fig.~\ref{fig:rt-5nm}\,(b). 
A fit of the resulting progression using the EEI correction, Eq.~\eqref{eq:qceeiT}, provides a screening factor of $\zeta=0.84$, 
which corresponds to weak screening and is in good agreement with charge carrier densities of $10^{12}$\,cm$^{-2}$,
extracted in Hall measurements (see Appendix~\ref{ap:hall}). 

Seemingly contradicting logarithmic dependences have also been discussed recently in the literature, \eg, in single crystal thin films of \BiSe~\cite{Wang2011, Liu2011}.
In these topological systems, a combination of two-dimensional quantum corrections of WAL and EEI has been used to describe the measured data.
However, we like to emphasize several aspects that are notably different in our data. 
First, the \BiSe samples show a multi-band signature in both the Hall as well the magnetoresistance data, but this signature has not been treated in this framework. 
In our case we find a clear single-band behavior instead. 
Secondly, the \BiSe magnetotransport data have been discussed for magnetic fields considerably smaller than the field that corresponds to the hypothetical WAL maximum.
We stress that our data cannot be explained over the full measured field range with this approach: 
the negative magnetoresistance for large fields, which is a main feature of quantum interference, is not captured by such an analysis. 
As far as we know, this issue is still not solved in more recent analyses, where the properties in higher magnetic fields are neglected~\cite{Lu2014, Sultana2018}. 

With the data evaluation suggested in this article, we consistently explain the magnetic field and temperature dependence jointly over the full accessible measured ranges:
Even when EEI is not pronounced in magnetotransport, it nonetheless plays a dominant role in the temperature dependence of the sheet resistance.
Especially in the case of weak screening, \ie, $\zeta \sim 1$, the magnetic field dependence in Eq.~\eqref{eq:qceei} vanishes, 
whereas the logarithmic temperature effect from EEI in Eq.~\eqref{eq:qceeiT} persists.
The magnetoresistance even shows an amplitude that exceeds the possible range of the EEI contribution independently of the screening effect by several orders of magnitude. 
Nevertheless, the screening has a strong impact on the temperature dependence of the sheet resistance.

\section{Summary}

In conclusion, we investigated electronic transport in thin films of \BPO down to low temperatures and up to high magnetic fields.
The large amplitude observed in the magnetoresistance as well as the distinct maximum cannot be explained within the scope of electron--electron interaction, 
but are both perfectly described by weak antilocalization with strong spin--orbit coupling as well as weak inelastic scattering. 

Our analysis had a clear course of action: 
first, we analyzed the temperature dependence of the weak antilocalization correction in our magnetoresistance data.
From there, we determined the sheet resistance by extrapolation to the respective zero magnetic field contribution of the magnetoresistance.
This zero magnetic field behavior, however, does not provide a simple logarithmic temperature dependence due to the finite inelastic field \Bi at zero temperature.
Only the data that are adjusted by subtraction of the weak antilocalization contribution display a plain logarithmic temperature dependence. 
This logarithmic dependence can in turn be interpreted with quantum corrections that result from electron--electron interaction for a weakly screened case, 
despite the negligible influence of interactions in the magnetoresistance.
Therefore, quantum corrections due to a mostly unscreened electron--electron interaction together with weak antilocalization provide a full quantitative understanding of our measured transport data. 

The lack of screening suppresses the effect of electron--electron interaction even further in the magnetoresistance (Eq.~\eqref{eq:qceei}), but its impact on the ground state of the electronic system is rather crucial. 
In the investigated \BPO films, the almost unscreened interaction is found to result in a positive contribution to the resistance for low temperatures---preempting a non-metallic state 
in spite of weak antilocalization. 

Further investigations are needed to study the screening in these systems.
Control of screening will allow to fine-tune the effect of electron--electron interaction to a scenario where~$\zeta \lesssim 1/2$, that is, 
where weak antilocalization predominates and the ground state is expected to be metallic. 
Further, we consider it worthwhile to investigate these specific properties of \BPO also in oxide heterostructures such as \BPO--\BBO~\cite{meir:2017}, 
where \BPO is in close vicinity to a charge-ordered insulator, or in interference with helical surface states of topological thin films. 

\begin{acknowledgments}
We thank Miriam Stroesser and Marcus Albrecht for helpful discussions and their support in the measurement of the magnetoresistance and Hall data.
This work was supported by the Deutsche Forschungsgemeinschaft (DFG, German Research Foundation) -- Grant number 107745057 -- TRR 80.
\end{acknowledgments}

\appendix

\section{Hall Measurement}
\label{ap:hall}

\begin{figure}
\centering
\includegraphics[width=\columnwidth]{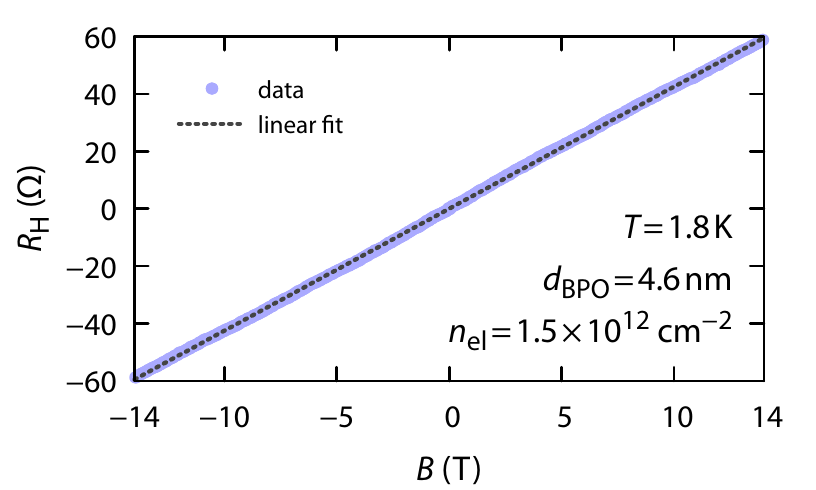}
\caption{Exemplary measurement of the Hall resistance (\RH) as function of magnetic field of a $\dBPO=\SI{4.6}{\nano\metre}$ thick \BPO thin film grown by pulsed laser deposition (PLD).
The Hall resistance~\RH depends linearly on the wide applied magnetic field strongly suggesting single-band behavior. 
The surface charge carrier density extracted from the linear fit (dashed line) calculates to $n_\mathrm{el}=\SI{1.5E12}{\per\square\centi\metre}$ at $\SI{1.8}{\kelvin}$.}
\label{fig:hall-bpo}
\end{figure}

Hall measurements in several samples of \BPO thin films show a linear behavior, suggesting a single type of charge carriers in our systems. 
An exemplary measurement for a $\SI{4.6}{\nano\metre}$ thick \BPO thin film up to \SI{14}{\tesla} is provided in Fig.~\ref{fig:hall-bpo}.

\section{Data Analysis of a 21.3\,nm \BPO Thin Film}
\label{ap:sample}

\begin{figure}
\centering
\includegraphics[width=\columnwidth]{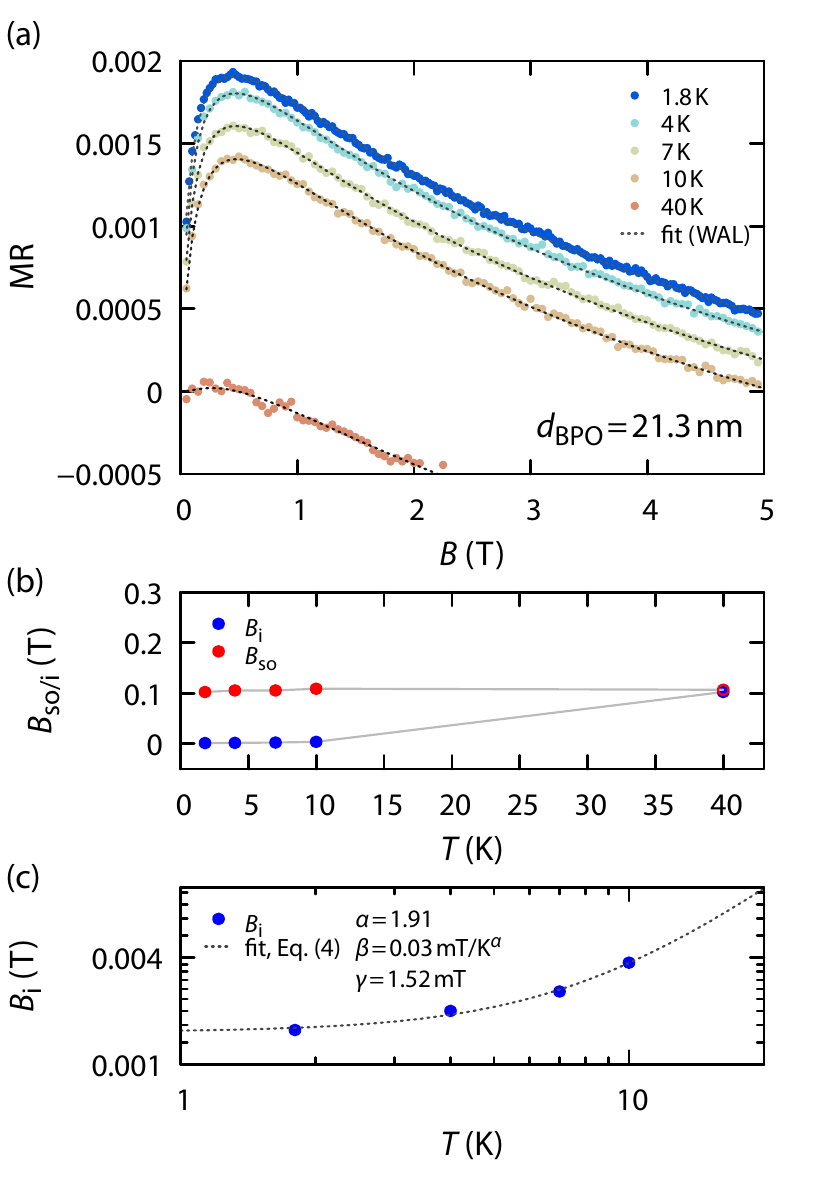}
\caption{Exemplary magnetoresistance (MR) data and fit parameter following Fig.~\ref{fig:wal5nm}, taken and retrieved from a $\SI{21.3}{\nano\metre}$ thick \BPO thin film
grown on a (001) oriented \STO substrate showing a maximum value of $\Bso=\SI{0.11}{\tesla}$.}
\label{fig:wal25nm}
\end{figure}

\begin{figure}
\centering
\includegraphics[width=\columnwidth]{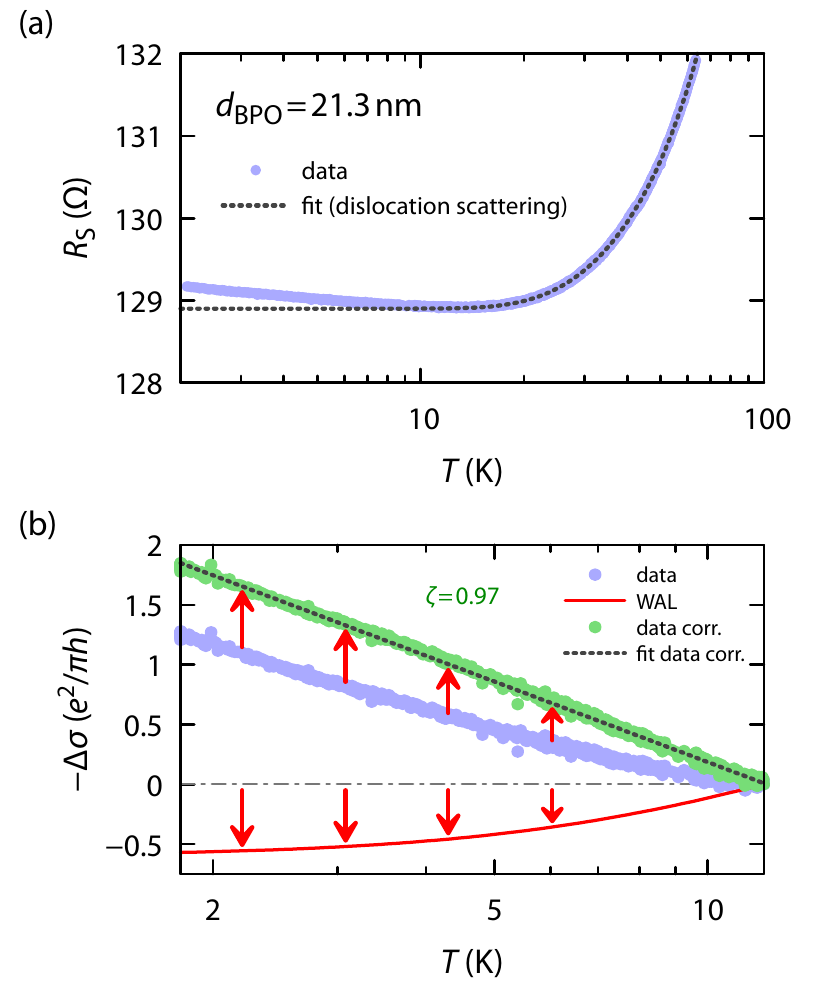}
\caption{(a) Exemplary observed dependence of the sheet resistance (R$_\text{S}$) on temperature retrieved for a $\dBPO=\SI{21.3}{\nano\metre}$ thick \BPO sample, following Fig.~\ref{fig:rt-5nm}. 
(b) The analysis follows the procedure described in the main text with chosen reference temperature $\Tref=\SI{12}{\kelvin}$ and reveals almost completely unscreened EEI modeled by $\zeta=0.97$.}
\label{fig:rt-21nm}
\end{figure}

The same data analysis procedure as shown in the main text also holds for other samples. 
We show exemplarily the results for a sample of a \BPO thin film of thickness $\SI{21.3}{\nano\metre}$. 
The spin--orbit field is smaller in the thicker sample as it has been in the thinner sample discussed in the main text, see Fig.~\ref{fig:wal25nm}. 
Again, the extrapolation of the weak antilocalization to zero fields allows to subtract the effect of quantum interference from the temperature dependent measurements, see Fig.~\ref{fig:rt-21nm}.
The remaining logarithmic increase of the resistance towards lower temperatures can be allocated unambiguously to EEI. 
In this case, the screening is nearly completely vanishing described by $\zeta=0.97$.

\end{document}